\documentclass{aa}

\usepackage{graphics}
\usepackage{epsfig}

\begin{document}

\title{Ionized iron K$\alpha$ lines in AGN X-ray spectra}

\author{Stefano Bianchi
	\and Giorgio Matt}

\offprints{Stefano Bianchi
		\email{bianchi@fis.uniroma3.it}}

\institute{Dipartimento di Fisica, Universit\`a degli Studi Roma Tre,
Via della Vasca Navale 84, I-00146 Roma, Italy}

\date{Received ; accepted }

\abstract{
The Equivalent Widths (EW) of the He-- and H--like iron lines produced in
photoionized, circumnuclear matter of AGN 
are calculated with respect to
both the reflected and the total continua. We found that the EWs with respect to
the total continuum may be as large as a few tens of eV, making them 
observable in bright Seyfert 1s 
by instruments on board $Chandra$ and XMM--$Newton$.
We apply our calculations to the XMM--$Newton$ spectrum of NGC~5506 and found a good
agreement with the data at the expense of a modest iron overabundance.
}
\maketitle

\section{Introduction}

The basic idea behind unification models for Seyfert galaxies
is that all objects
are intrinsically the same and all observed different features are simply
the result of the orientation of the system with respect to the line
of sight.

In this scenario, emission from circumnuclear regions is much easier to observe
in Seyfert 2s (and in particular in Compton--thick ones),
where the nuclear radiation is obscured, than in
unobscured Seyfert 1s in which these components are heavily diluted by the
photons from the nucleus.
Indeed, such regions are commonly observed in Seyfert 2s (e.g. Matt 2001
for a review), and found to be of at least
two types: a cold gas with high column density which partially or
completely obscures the nuclear radiation (the `torus') and an ionized,
optically thin
medium that scatters it into the line of sight. Both these materials produce
a reflected continuum and narrow iron K$\alpha$
lines (e.g. Matt et al. 1996, hereinafter MBF96).

If these emitting regions are also present in the environment of Seyfert 1s, as
expected in unification schemes,
we should in principle be able to observe these lines as well, although
with a much smaller EW due to the dilution by the direct nuclear continuum. The
limited sensitivity and energy resolution of past X--ray missions prevented,
until a couple of years ago, the unambiguous detection of these features in
unobscured sources. The situation has been radically changed with
the launch of $Chandra$ and XMM--$Newton$. Indeed,
a narrow neutral iron line have been found to be fairly common
in Seyfert 1s spectra  (e.g.
Kaspi et al. 2001; Yaqoob et al. 2001; Weaver 2002; Yaqoob et al
2002 for $Chandra$ results; Reeves
et al. 2001; Pounds et al. 2001; Gondoin et al. 2001; Matt et al. 2001
for XMM--$Newton$ findings). On the other hand,
evidence for ionized iron K$\alpha$ lines at 6.7 and 6.96 keV from
distant matter have been reported so far in only one
unobscured (at the lines energies) Seyfert, namely NGC~5506 (Matt et al. 2001).

In this paper, we calculate the expected equivalent widths for the
two ionized lines with respect to both reflected and total continua.
Details of the calculations and results will be presented in Sects. 2 and 3,
respectively, and the case of NGC~5506 discussed in Sect. 4.

\section{The model}\label{model}

The line emitting material is assumed to be
optically thin to Thomson scattering and in photoionization
equilibrium, as observed e.g. in the
`hot reflector' in NGC~1068 (see Guainazzi et al.
1999; Bianchi et al. 2001).
In these conditions, both recombination and resonantly
scattered K$\alpha$ lines from H-- and He--like iron ions can be produced.

Resonantly scattered  K$\alpha$ lines consists of radiative
de--excitation following excitation from continuum photons.
The radiative
channel has a probability of occurrence which, for Fe~{\sc xxv} and
Fe~{\sc xxvi}, is unity as Auger de--excitation is of course impossible.
The angular dependence of the emission depends on the transition involved
(Chandrasekhar 1960).
In the column density intervals explored in this paper, resonant lines
are optically thick.
Therefore, the first resonant scattering occurs preferentially close to
the illuminated surface (this effect of course becoming more and more
important as the column density of the matter increases). As there are
not competing processes apart from Compton scattering (which has a much smaller
cross section), the line photons eventually escape from the matter,
but with a larger probability to escape from the illuminated  surface
(reflection case, see Fig.~1) rather than from the opposite boundary
(transmission case). Line transfer in the matter is treated here
by means of Monte Carlo simulations.

Recombination K$\alpha$ lines occur after a photoionization of a
K shell electron. Similarly to the case of fluorescence, it is customary to
describe the process by introducing an effective fluorescent yield $Y$, which is the
probability that the recombination cascade does include the K$\alpha$
transition (Krolik \& Kallman 1987; MBF96).
Recombination occurs almost homogeneously in the matter,
as it follows photoabsorption, which is optically thin in the column density
ranges considered here. Of course, K$\alpha$
line photons produced by recombination
may in turn be resonantly scattered before escaping, but because
of the uniform source function there is no preferred escaping
surface. This process can be therefore treated in the optically thin
approximation, thus analytically.

Both processes and their treatment
are described in detail in MBF96, to which the reader is deferred.

Ionization fractions for Fe~{\sc xxv} and
Fe~{\sc xxvi} have been calculated with the well known
public photoionization code
\textsc{cloudy}\footnote{http://www.pa.uky.edu/\~{}gary/cloudy/}
(v94.00; Ferland 2000), as a function of the
ionization parameter $U$, defined as:

\begin{equation}
U=\frac {\int _{\nu _{R}}^{\infty }\frac{L_{\nu }}{h\nu} d\nu}
{4\pi r^{2}cn_{e}}
\end{equation}

\noindent where $c$ is the speed of light, $r$ the distance of the gas from the
illuminating source, $n_{e}$ its density and $\nu_{R}$ the frequency
corresponding to 1 Rydberg.
This, however, is not the best choice when investigating highly ionized
iron atoms, as the result depends dramatically on the spectral shape
of the photoionizing continuum. Assuming a power law shape, a small
change in the spectral index (for the same $U$)
may change significantly the number of
X--ray photons (which are the most relevant in this case).
We decided to use  instead a ``X--ray"
ionization parameter, defined over the 2--10 keV energy range:

\begin{equation}
U_{x}=\frac {\int _{2}^{10}\frac{L_{\nu }}{h\nu} d\nu} {4\pi r^{2}cn_{e}}
\end{equation}

\noindent where the integration limits are the frequencies correspondent to 2
and 10 keV.
The conversion factor between the two ionization parameters is
clearly a function of the power index of the continuum:

\begin{equation}
\frac {U_{x}} {U} (\gamma) =\frac {\int _{2}^{10}\frac{L_{\nu }}{h\nu} d\nu}
{\int _{\nu _{R}}^{\infty }\frac{L_{\nu }}{h\nu} d\nu} = -
\frac {10^{1-\gamma} - 2^{1-\gamma}} {E_{R}^{1-\gamma}}
\end{equation}

\noindent where $\gamma$ is the photon index of the powerlaw and
$E_{R}=13.6\times10^{-3}$ keV.
Adopting this new definition of the ionization parameter, the results are
almost independent of $\gamma$: the residual effects will be presented
at the end of the next section, together with those due to changing
the temperature and density of the gas.

\begin{figure}
\resizebox{\hsize}{!}{\includegraphics{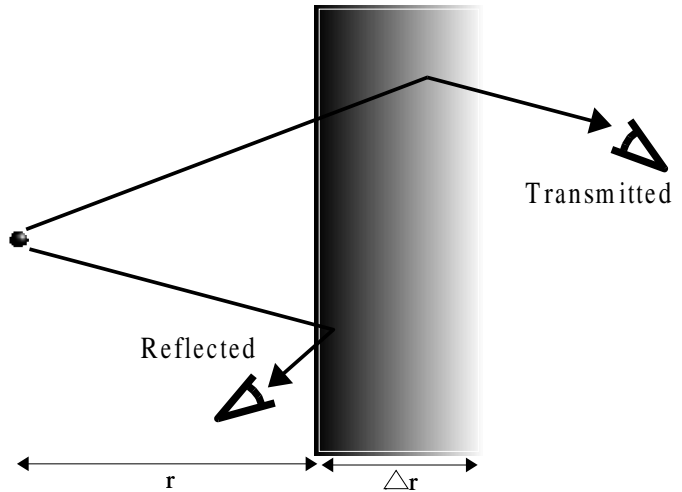}}
\caption{The adopted geometry.}
\label{geometry}
\end{figure}

The sketch in Fig.\ref{geometry} shows the geometry adopted in our
calculations. We assumed $\Delta r$=$r$ and isotropic
illumination within a 30$^{\circ}$ cone.
As said above, the equivalent widths of the resonant scattered lines
are different when transmitted or reflected spectra are considered.
Therefore, we will present plots for both cases.

\section{Results}

\begin{figure*}[htb]
\begin{minipage}[t]{87mm}
\epsfig{file=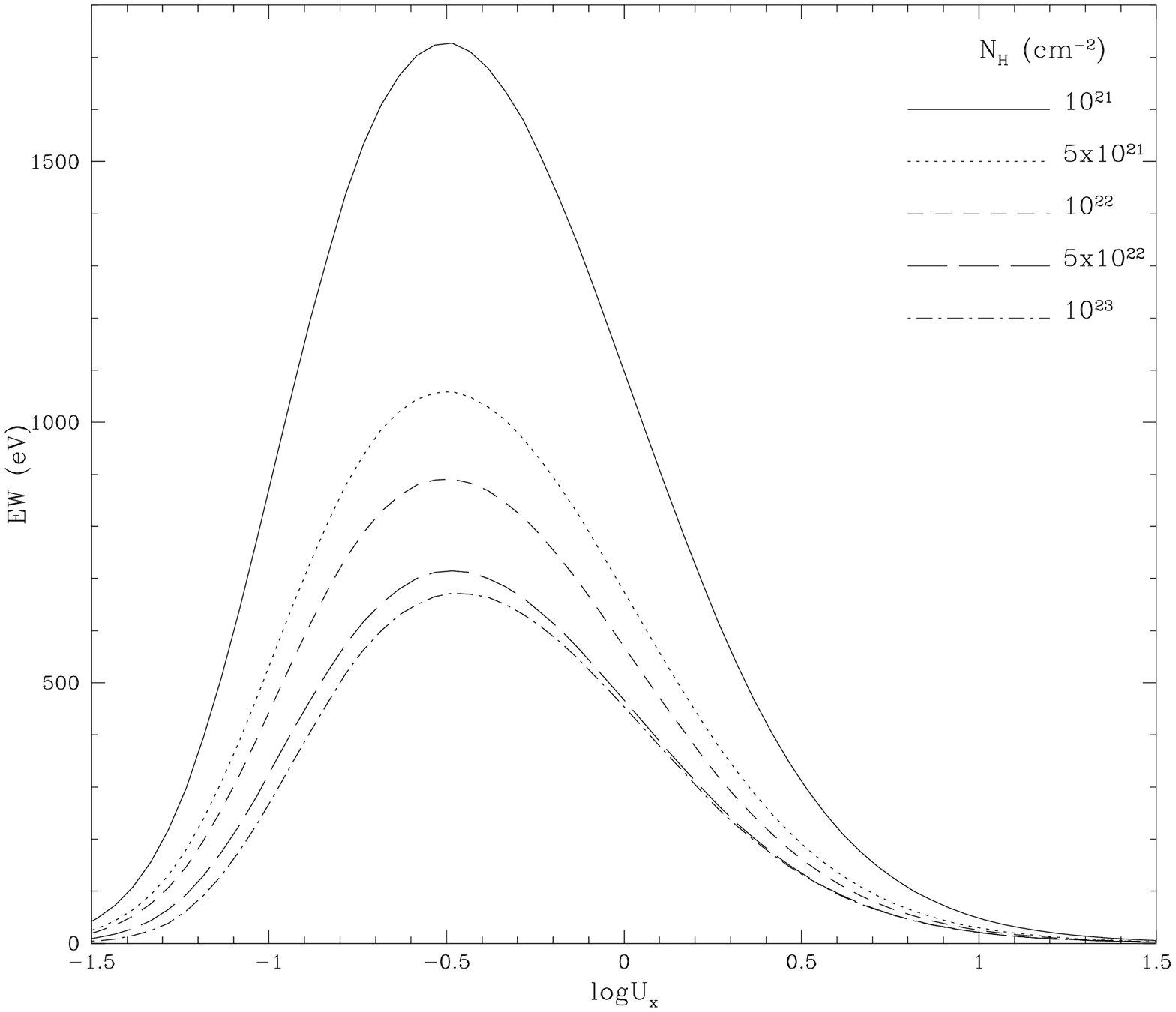,height=90mm}
\caption{Fe {\sc xxv} EWs against transmitted continuum only as a function of
$\log U_{x}$ and column density.}
\label{XXV}
\end{minipage}
\hspace{\fill}
\begin{minipage}[t]{87mm}
\epsfig{file=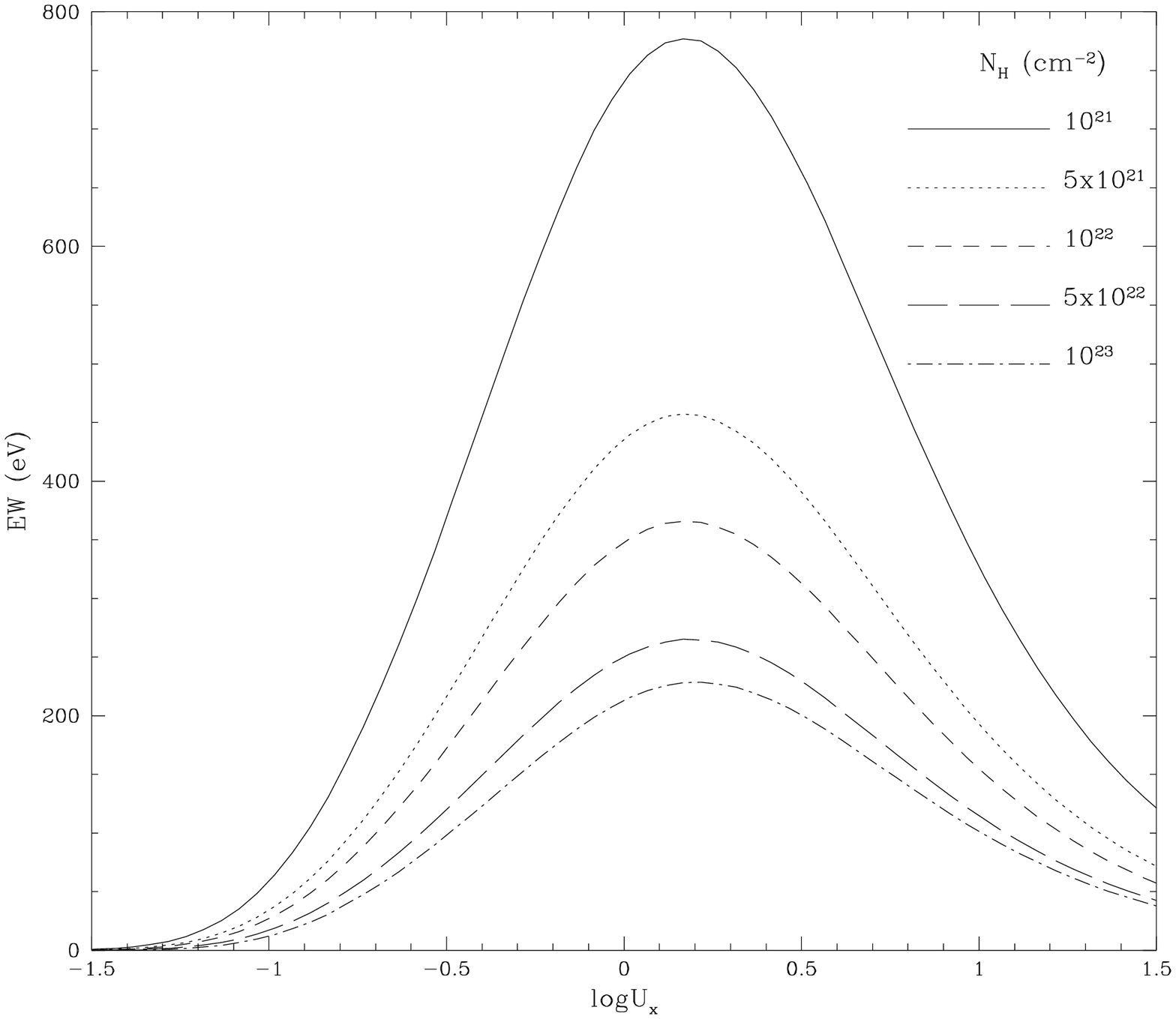,height=90mm}
\caption{Fe {\sc xxvi} EWs against transmitted continuum only as a function of
$\log U_{x}$ and column density.}
\label{XXVI}
\end{minipage}
\end{figure*}

\begin{figure*}[htb]
\begin{minipage}[t]{87mm}
\epsfig{file=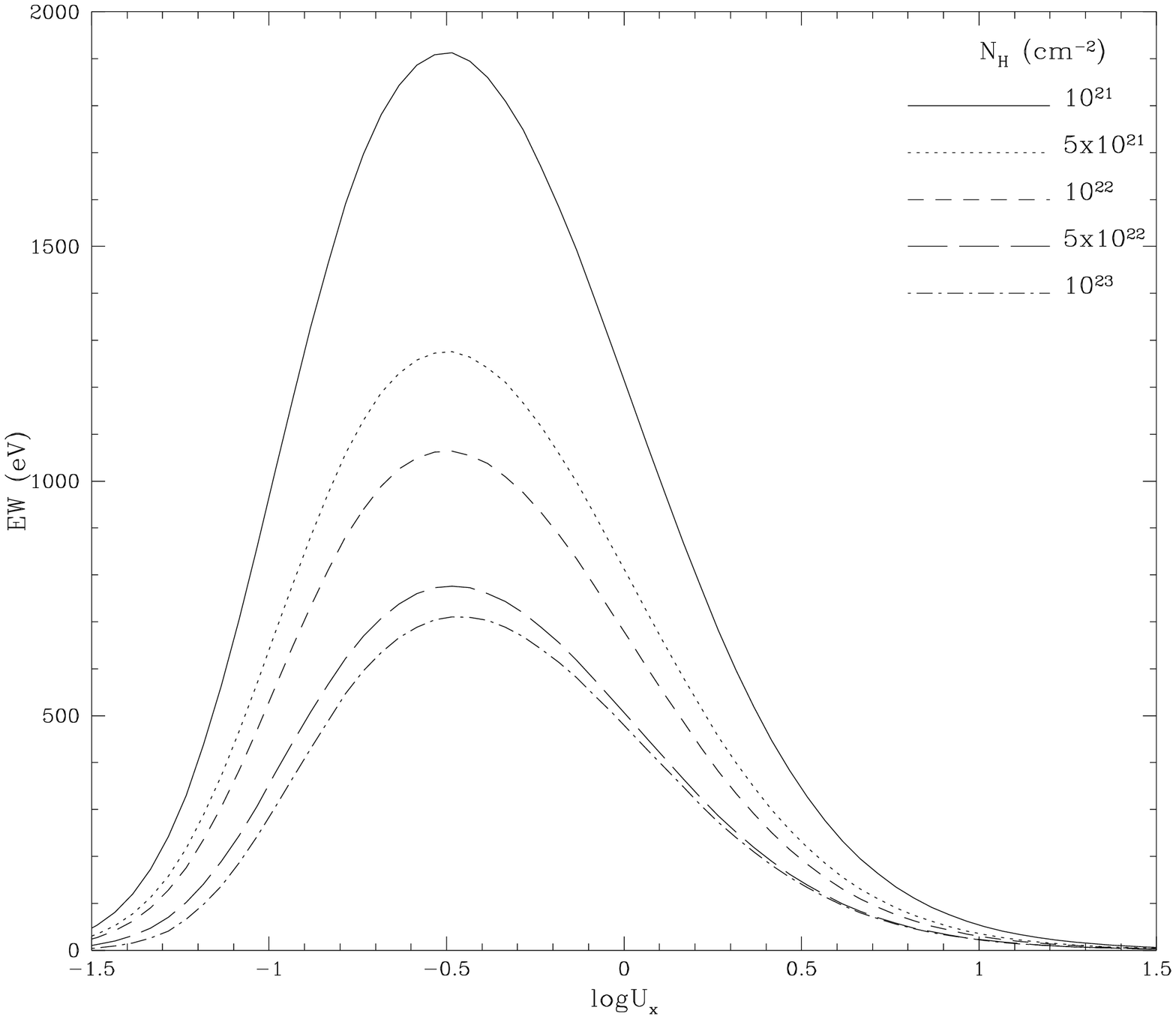,height=90mm}
\caption{Fe {\sc xxv} EWs against reflected continuum only as a function of
$\log U_{x}$ and column density.}
\label{XXVR}
\end{minipage}
\hspace{\fill}
\begin{minipage}[t]{87mm}
\epsfig{file=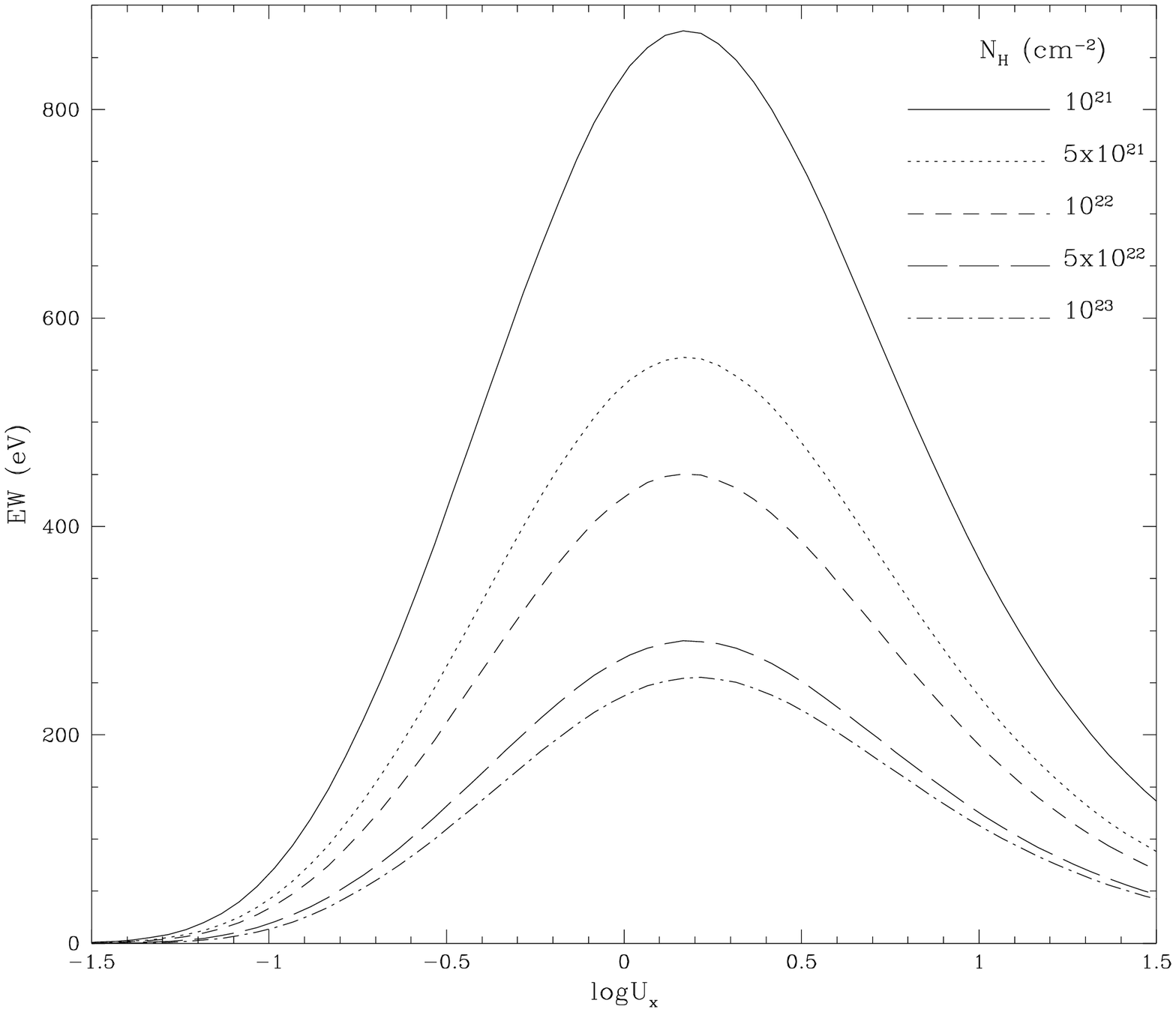,height=90mm}
\caption{Fe {\sc xxvi} EWs against reflected continuum only as a function of
$\log U_{x}$ and column density.}
\label{XXVIR}
\end{minipage}
\end{figure*}

In Figs.~\ref{XXV}, \ref{XXVI}, \ref{XXVR} and \ref{XXVIR}
the equivalent widths
of Fe~{\sc xxv} and Fe~{\sc xxvi} K$\alpha$
lines (recombination plus resonantly scattered; hereinafter we will
always consider the sum of the two contributions) against the transmitted (or
reflected) continuum only are plotted as a function of $\log U_{x}$ and for
different values of the column densities of the line emitting gas.
The results are obtained with $\gamma=2.0$,
$T=10^{6}$ K, $n_{e}=10^{6}\,cm^{-2}$.
Plots in Figs.~\ref{XXV} and \ref{XXVI}
were obtained combining the results in figure
4, 5 and 6 of MBF96 (where EWs of recombination and resonant lines where
presented assuming ionization
fractions equal to 1 for the interested ion) with the ionization structures
calculated by \textsc{cloudy} as a function of $U_{x}$. The iron abundance
is assumed equal to the solar one (as for Morrison \& McCammon 1983).
As described in MBF96,
for small column densities the main contribution comes from the resonant
lines, while at large column densities  recombination
dominates, because photoionization is still optically
thin while resonant scattering is thick, and the line production for the
latter process consequently saturates.
Plots in Figs.~\ref{XXVR} and \ref{XXVIR}
have been obtained re--running the Monte Carlo code described in
MBF96 for a reflecting, rather than transmitting, material.
Because the two cases differs only for the contribution of the resonant
lines, the main differences occur for small column densities.

\begin{figure*}
\begin{minipage}[t]{87mm}
\epsfig{file=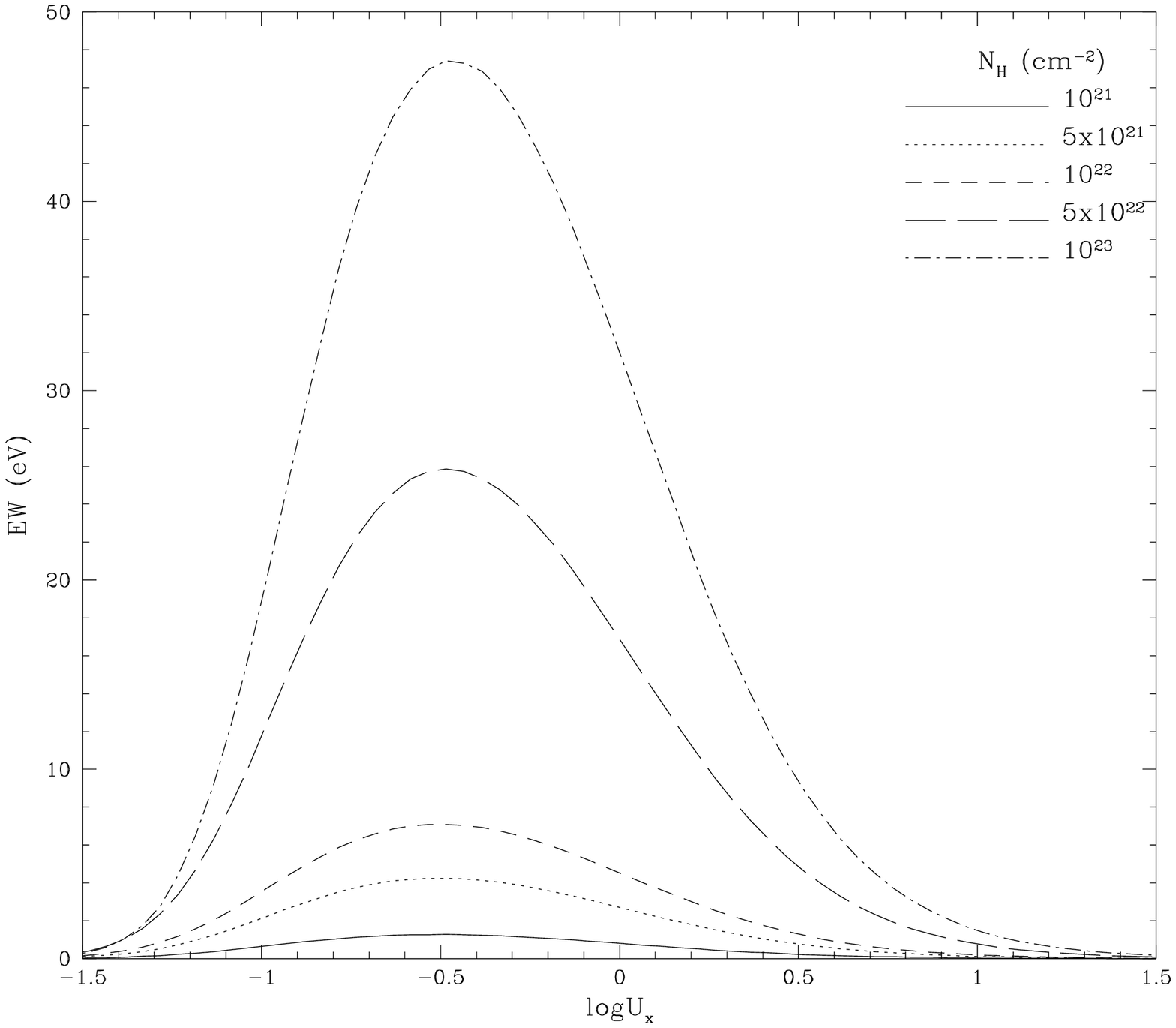, height=90mm}
\caption{Reflected Fe {\sc xxv}
EWs against total continuum as a function of $\log
U_{x}$ and column density (f=1: see text for details).}
\label{XXVtotR}
\end{minipage}
\hspace{\fill}
\begin{minipage}[t]{87mm}
\epsfig{file=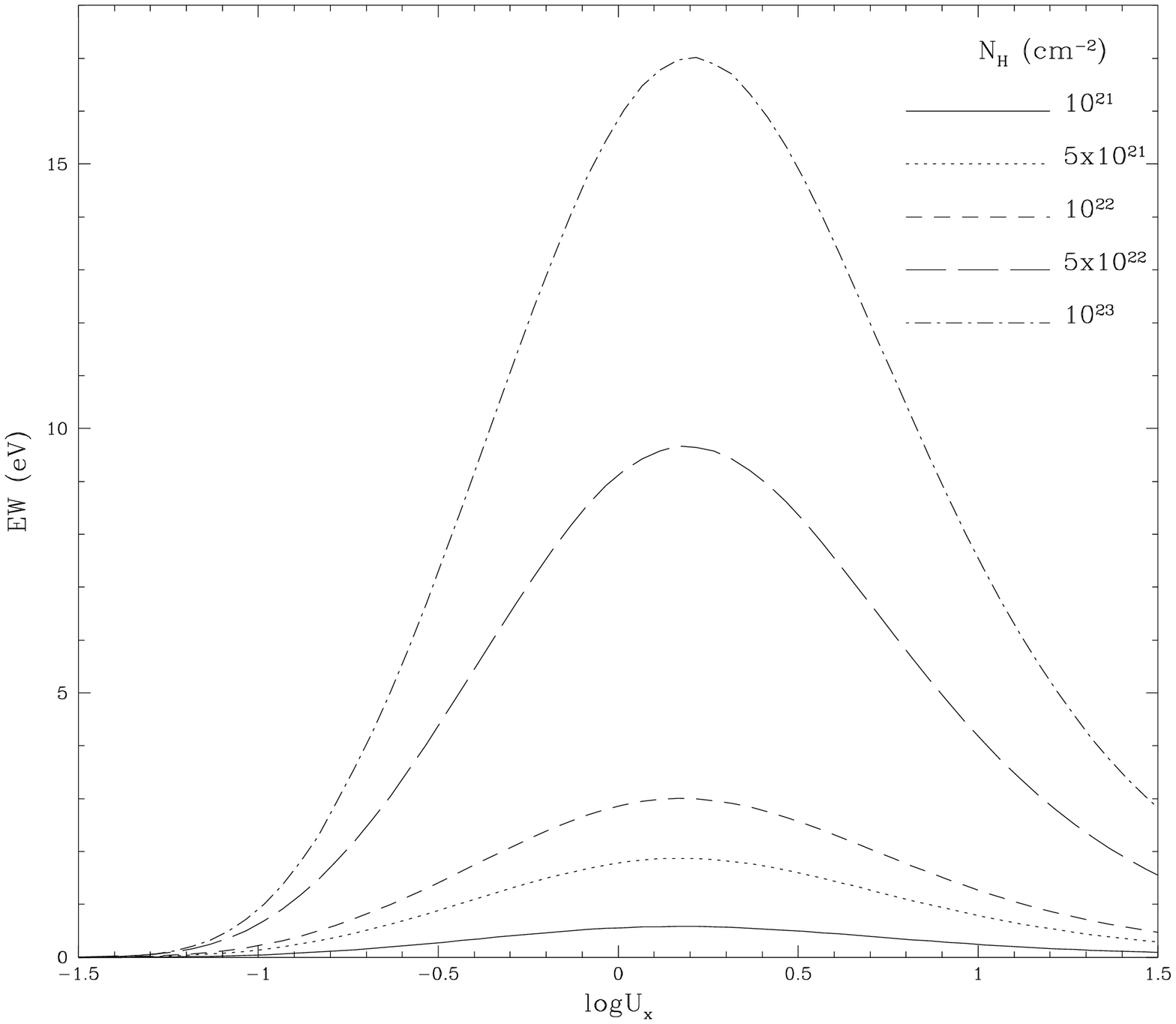,height=90mm}
\caption{Reflected Fe {\sc xxvi}
 EWs against total continuum as a function of
$\log U_{x}$ and column density (f=1: see text for details).}
\label{XXVItotR}
\end{minipage}
\end{figure*}

In Figs.~\ref{XXVtotR} and \ref{XXVItotR}
the EWs, obtained for the reflecting matter,
 are instead calculated against the total continuum.
(The corresponding
plots for the transmitted flux are not shown because differences are
appreciable only for small column densities, where the
EWs are very small.)
The relation between the EWs with respect to reflected and
total continua is easy to calculate. In the Compton--thin regime,
the reflected flux is given by:

\begin{equation}
F_{r}\simeq f\cdot \tau \cdot F_{t}
\label{FrFt}
\end{equation}

\noindent where $F_{t}$ is the illuminating flux and
$f$ a geometrical factor (basically the solid angle subtended by the
illuminated matter);
$\tau= \frac {N_{H}} {1.5\times 10^{24}}$ is the Thomson optical
depth. Therefore, the equivalent widths are:

\begin{equation}
EW_{t} = EW_{r} \cdot \frac {F_{r}} {F_{t} + F_{r}} \simeq   EW_{r}
\cdot \frac {F_{r}} {F_{t}} \simeq EW_{r} \cdot f\cdot \tau
\end{equation}
\noindent Figs.~\ref{XXVtotR} and \ref{XXVItotR} have been obtained
assuming $f$=1.

\begin{figure*}
\begin{minipage}[t]{87mm}
\epsfig{file=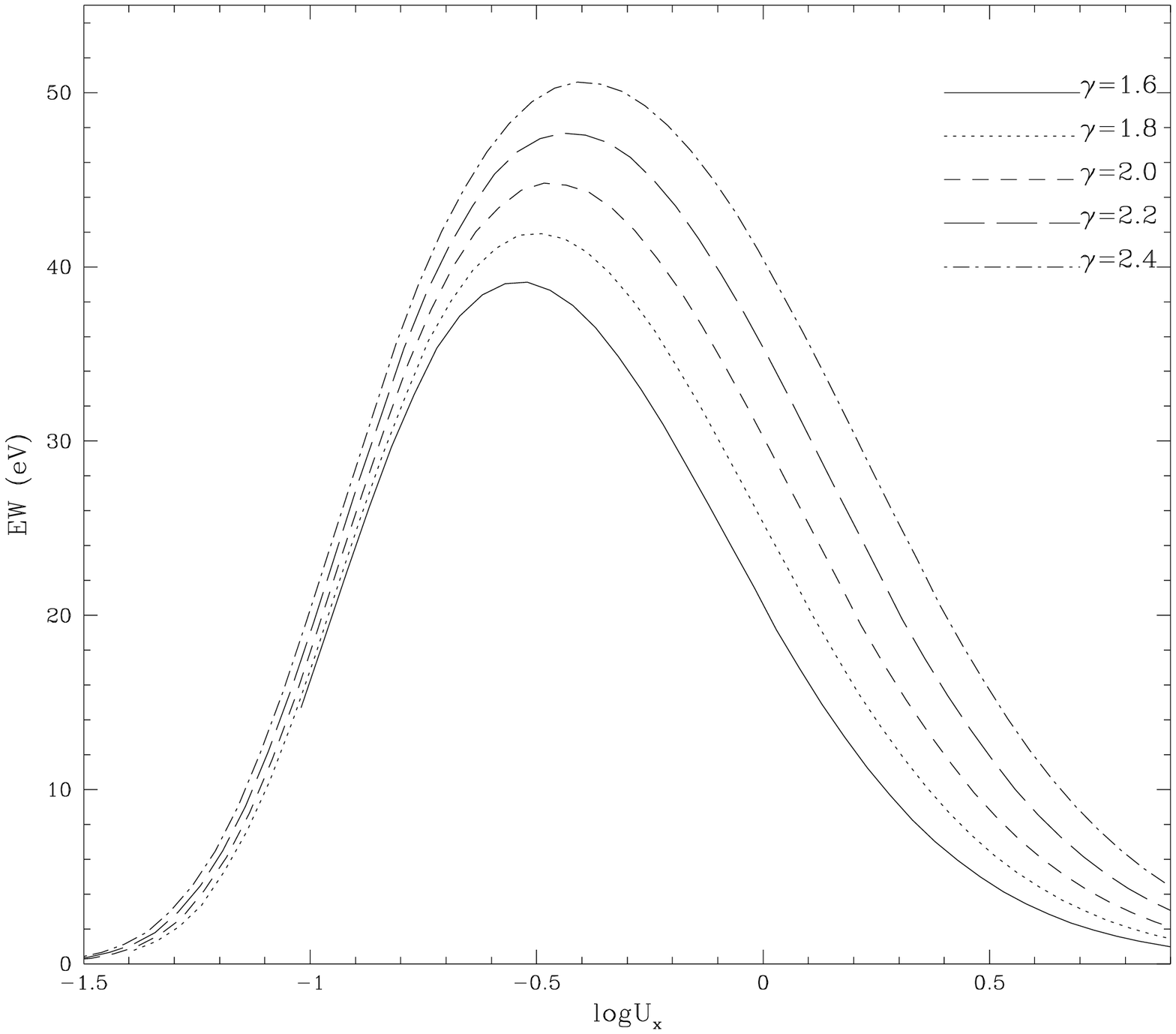,height=90mm}
\caption{Fe {\sc xxv} EWs against total continuum for different values
of $\gamma$. Column density is $10^{23}$~cm$^{-2}$.}
\label{XXVgamma}
\end{minipage}
\hspace{\fill}
\begin{minipage}[t]{87mm}
\epsfig{file=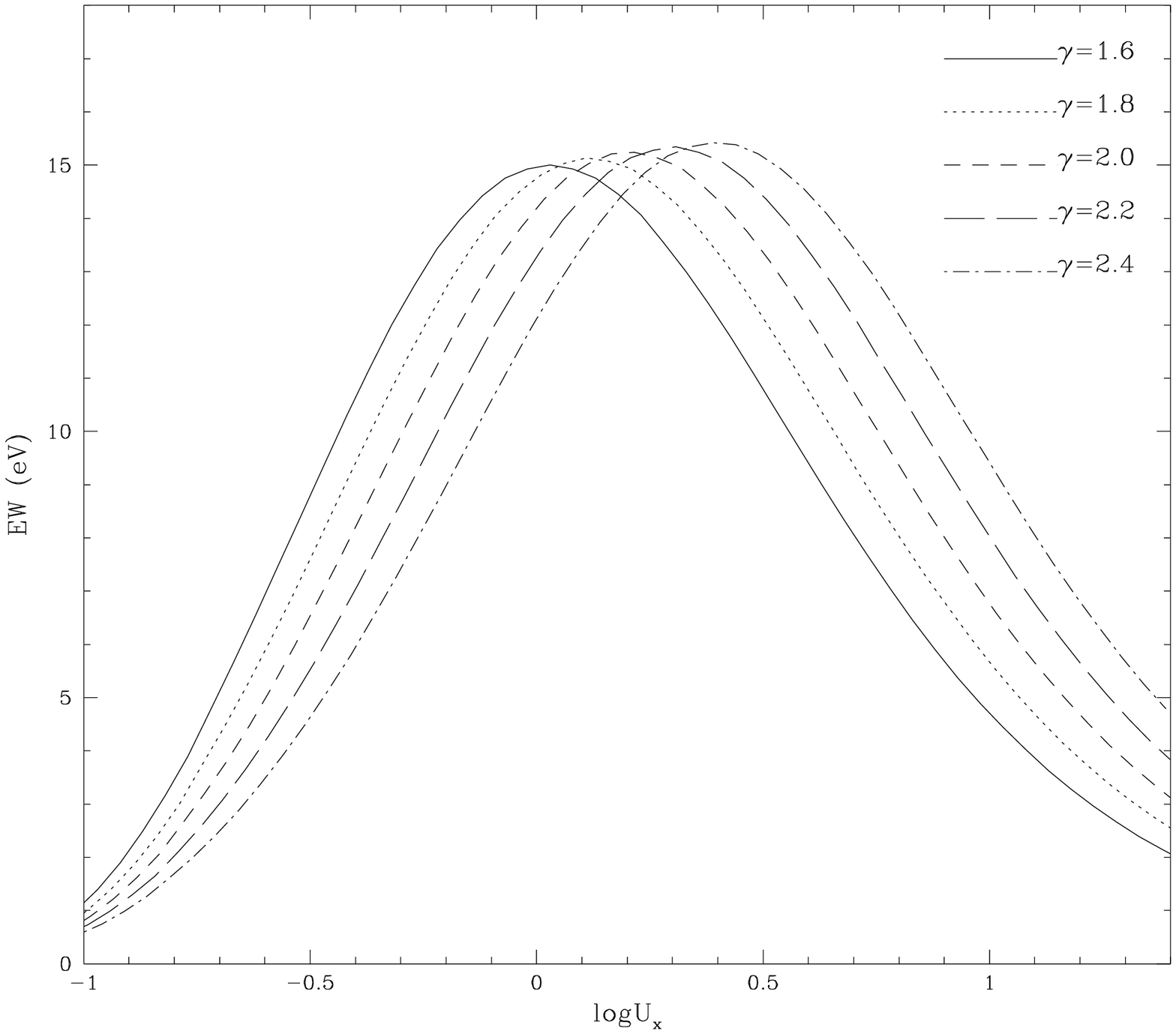,height=90mm}
\caption{Fe {\sc xxvi} EWs against total continuum for different values
of $\gamma$. Column density is $10^{23}$~cm$^{-2}$.}
\label{XXVIgamma}
\end{minipage}
\end{figure*}

Finally, we illustrate the effects of changing the value of
$\gamma$, $T$ and $n_{e}$.
Fig.~\ref{XXVgamma} shows the equivalent widths
of Fe~{\sc xxv} lines against the
total continuum for a column density of
$10^{23}$~cm$^{-2}$ and different values of the photon index of the
incident continuum.
Clearly, the choice of $U_{x}$ instead of $U$ keeps the differences small:
the maximum of the function occurs almost at the same value of
$U_{x}$ for the various indices,
while it would have been significantly
shifted if plotted against $U$. There is however
a difference in the value of the maximum, which
increases with $\gamma$.

The opposite occurs for Fe~{\sc xxvi}
(Fig.~\ref{XXVIgamma}), with a shift in
the value of $U_{x}$ for which the function has the maximum value, but
not appreciable differences in the values of this maximum.

\begin{figure*}
\begin{minipage}[t]{87mm}
\epsfig{file=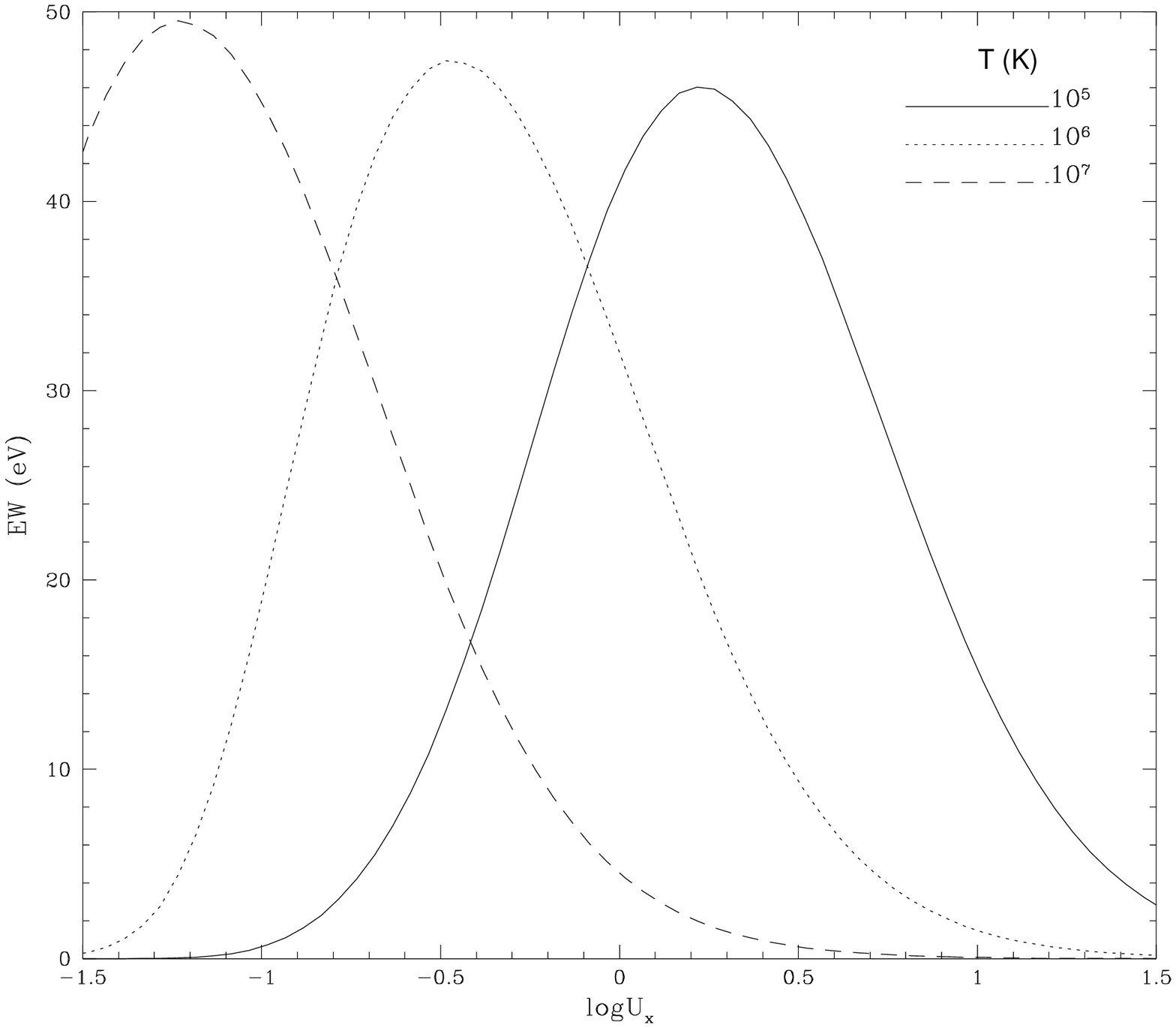,height=90mm}
\caption{Fe {\sc xxv} EWs against total continuum for different values
of $T$. Column density is $10^{23}$~cm$^{-2}$.}
\label{XXVtemp}
\end{minipage}
\hspace{\fill}
\begin{minipage}[t]{87mm}
\epsfig{file=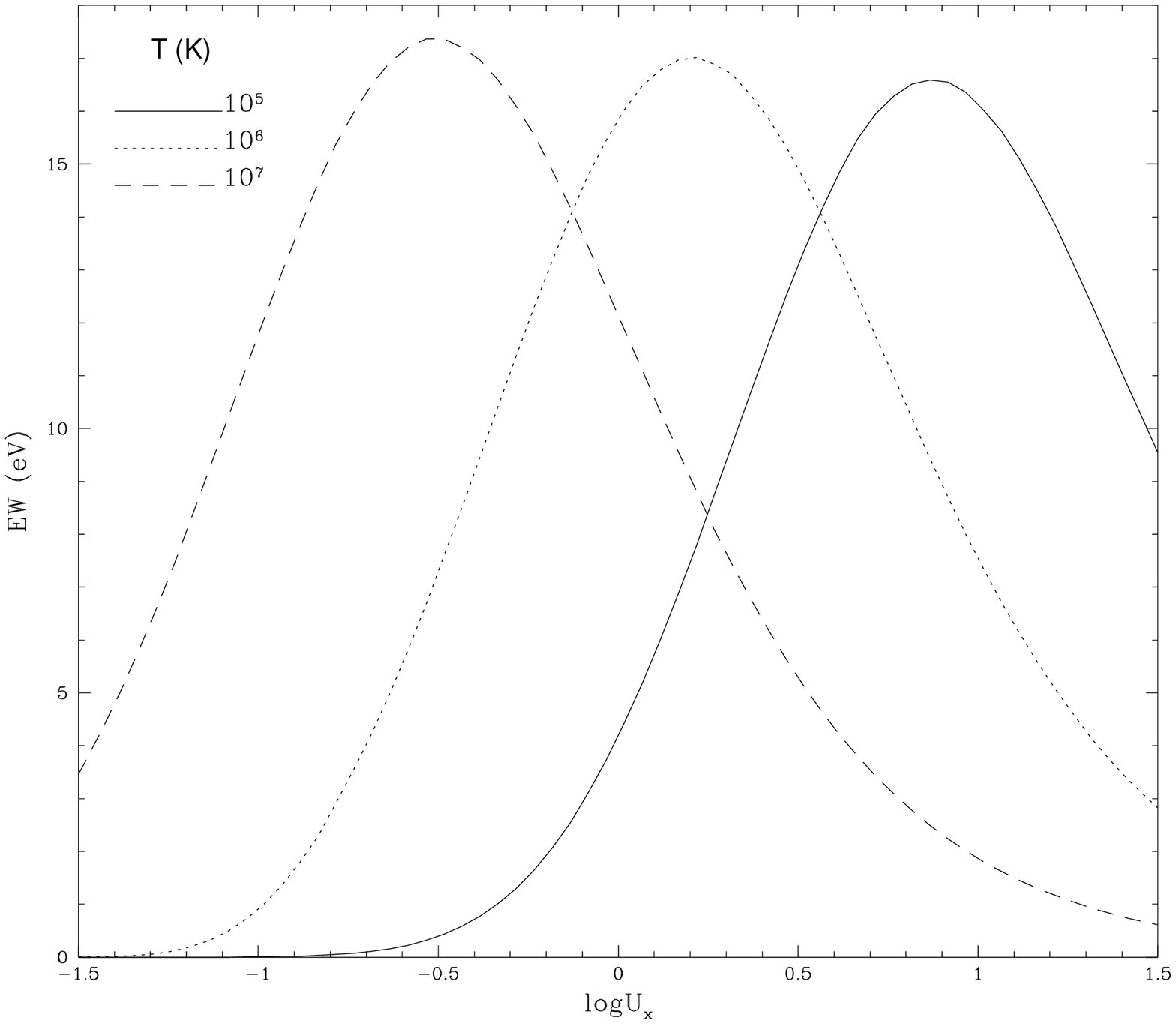,height=90mm}
\caption{Fe {\sc xxvi} EWs against total continuum for different values
of $T$. Column density is $10^{23}$~cm$^{-2}$.}
\label{XXVItemp}
\end{minipage}
\end{figure*}

The effect of changing the gas temperature is shown in
Figs.~\ref{XXVtemp} and
\ref{XXVItemp}. The overall result is, as expected, that the colder is the gas
the higher is the ionization parameter required to reach the peak in the
ionization fraction of both the iron ions. On the other hand, there are
 no significant
changes in the maximum values of the equivalent widths of the lines.

A change of the gas density over a wide range ($10^4 - 10^8$
cm$^{-3}$) does not instead produce any appreciable
variations in our results. This is
expected, because the ionization parameter is a quantity that describes the
ionization structure of a gas under a combination of density, distance and
luminosity of the incident radiation. Changing one of these quantities
keeping the ionization parameter fixed ends up only in a rescale of the other
ones, without changing the ionization structure of the gas surface.

\section{An example: NGC~5506}

Let us now compare our calculations with the only case in which
the presence of ionized iron K$\alpha$ lines from circumnuclear matter
have been claimed in an unobscured (at the iron line energies) source,
namely NGC~5506 (Matt et al. 2001).

NGC~5506 is a bright Narrow Emission Line Galaxy, optically classified
as an intermediate Seyfert with a highly reddened BLR (Veron-Cetty \& Veron
2001). Several X-ray observations clearly indicate the presence of a hidden
Seyfert 1 nucleus (Matt et al. 2001 and references therein). Its iron
K$\alpha$ line has been claimed to be complex by {\it ASCA} (Wang et al. 1999)
and {\it RXTE} (Lamer et al. 2000) observations. {\it BeppoSAX} found an iron
line centroid energy definitely higher than 6.4 keV (Perola et al. 2002).

XMM--{\it Newton} data definitely confirm that the iron line is
complex, with at least two components clearly present:
one narrow, corresponding to neutral iron, the other broad
and consistent with ionized iron.

Even if a fit of the ionized, broad component with a disc relativistic
line is as good as
that with a blend of He-- and H--like unresolved lines, the second
solution is preferable
from a physical point of view, as argued by Matt et al. (2001).
If this is, indeed, the case, these two lines are likely to be
produced in a photoionized matter as the one described in this paper.
The observed equivalent widths are $40\pm16$ eV and $32\pm15$ eV for
Fe~{\sc xxv} and Fe~{\sc xxvi}, respectively.

Moreover, NGC~5506 presents a soft excess whose normalization
is about 2\% of the primary component:
recalling equation~(\ref{FrFt}) this could be the result of reflection
from  a gas with column densities of $3\times10^{22}/f$ cm$^{-2}$.
Let us now assume that the two ionized lines and the
soft X-ray emission come from one and the same matter. Let us also
assume $f$=1 (a smaller value of $f$ would implies a larger value of
the column density to produce the observed flux of the reflected
continuum: the two parameters have the same effect on the line
EWs, and in the first approximation we can assume that the EWs are
almost independent of their ratio). The observed EWs can be reproduced
provided that $\log(U_x)\sim0\div0.5$ ($T=10^{6} K$) and the iron is
overabundant by a factor 2$\div$3. (A different choice of the gas temperature
would require a different value of $\log(U_x)$, but not of the iron abundance,
see Figs.~10 and 11).
This solution is of course valid in the
stationary case. If the source flux was larger in the past than during the
XMM--$Newton$ observation, then the requirement on the iron overabundant will be
correspondingly reduced; it is worth noting that the flux during the
XMM--$Newton$ observation was about 30\% lower than the average flux
during the longer contemporaneous $BeppoSAX$ observation.

Finally, combining the derived value of the ionization parameter with the
luminosity of the object ($L_{2-10}=8.5\times10^{42}$ erg cm$^{2}$ s$^{-1}$),
its spectral shape ($\Gamma \simeq 2$)
and the inferred column density of the gas, we can write two equations
linking the above, observed parameters with
the density $n_{e}$, the distance $r$ (from the nucleus) and the depth
$\Delta r$ of the gas (see Fig.~\ref{geometry}):

\begin{equation}
n_{e}r^{2}=\frac {\int _{2}^{10}\frac{L_{\nu }}{h\nu} d\nu} {4\pi c U_{x}}\\
n_{e}\Delta r = N_{H}
\end{equation}

\noindent where $r=\Delta r$ in our model (see Sect. \ref{model}). If
we use the value of $U_x$ found before to reproduce the observed lines for
$T=10^{6}$ K, we get the following estimates for $r$ and $n_{e}$:

\begin{equation}
n_{e}\simeq2.5\times10^5/f ~~~{\rm cm}^{-3}~;\\
r\simeq0.04\cdot f  ~~~{\rm pc}
\end{equation}

These values change if we use the ionization parameters which best reproduce the
observed EWs in the case of $T=10^{5}$ K and $T=10^{7}$ K (see
Figs.~\ref{XXVtemp} and \ref{XXVItemp}), being respectively:
[$n_{e}\simeq8\times10^5/f ~{\rm cm}^{-3}$; $r\simeq0.01\cdot f ~{\rm pc}$] and
[$n_{e}\simeq8\times10^4/f ~{\rm cm}^{-3}$; $r\simeq0.1\cdot f ~{\rm pc}$].

This distance scale is slightly larger than that typical of the BLR
(e.g. Peterson 1997). On the other hand, it is definitely smaller than the
inner radius of the absorbing torus, which is of the order of parsecs (e.g.
Bianchi et al. 2001). Interestingly, our estimated distance is consistent with
the (poorly constrained) distance scale of warm absorbers (e.g. Otani et
al. 1996; Netzer et al. 2002), but it is worth remarking that the ionization
parameter required in NGC~5506 is clearly higher than that of warm absorbers.

\section{Conclusions}

We have calculated the equivalent widths of H-- and He--like iron K$\alpha$
lines from photoionized, Compton--thin matter in Seyfert galaxies,
using results from MBF96 and ionization fractions estimated with
\textsc{cloudy}.
We have applied our calculations to NGC~5506, whose XMM-{\it
Newton} spectrum shows a complex iron K$\alpha$ line system, possibly
composed by a neutral line and a blend of two ionized lines as those produced
in our model. We found that, if this interpretation is correct, the
ionized lines observed in this source can be reproduced assuming $\log
U_{x}\sim0\div0.5$ (if $T=10^6$ K) and an iron overabundance of $2\div3$.

At a given gas temperature, the range
of ionization parameters which produces EWs as observed in NGC~5506 is
rather narrow; moreover, NGC~5506 is one of the brightest
AGN in the X--ray sky. We then
expect that, unless for some unknown reason the ionization parameter
tends to assume values in this narrow range, in most cases these lines will be
below detectability even with XMM--$Newton$ and $Chandra$, and therefore that
NGC~5506 may be the exception rather than the rule. It is interesting to note
that very similar iron line complexes have been observed so far by XMM--{\it
Newton} in only two other sources, Mrk~205 (Reeves et al. 2001) and Mrk~509
(Pounds et al. 2001) where, as discussed in those papers,
the origin of the ionized lines is likely to be the accretion disc.

\begin{acknowledgements}
We thank G. Cesare Perola for useful comments and suggestions, and the
anonymous referee for his help in improving the clarity of the paper. Financial
support from ASI and MIUR (under grant {\sc cofin-00-02-36}) is acknowledged.
\end{acknowledgements}

\end{document}